\documentclass[aps,prd,superscriptaddress,showpacs,preprint]{revtex4}
\usepackage{graphicx, bm}
\usepackage[usenames]{color}

\begin{document}
\draft
\title{Improved bounds on the dipole moments of the tau-neutrino at high-energy $\gamma^* e^-$ and $\gamma^* \gamma^*$ collisions: ILC and CLIC}

\author{ A. Guti\'errez-Rodr\'{\i}guez\footnote{alexgu@fisica.uaz.edu.mx}}
\affiliation{\small Facultad de F\'{\i}sica, Universidad Aut\'onoma de Zacatecas\\
         Apartado Postal C-580, 98060 Zacatecas, M\'exico.\\}

\author{M. Koksal\footnote{mkoksal@cumhuriyet.edu.tr}}
\affiliation{\small Deparment of Physics, Cumhuriyet University, 58140, Sivas, Turkey.\\}

\author{A. A. Billur\footnote{abillur@cumhuriyet.edu.tr}}
\affiliation{\small Deparment of Physics, Cumhuriyet University, 58140, Sivas, Turkey.\\}

\date{\today}

\begin{abstract}

In this work we study the potential of the processes $e^+e^- \to e^+\gamma^* e^- \to e^+\tau \bar\nu_\tau \nu_e $
and $e^+e^- \to e^+\gamma^* \gamma^* e^- \to e^+\nu_\tau \bar\nu_\tau e^- $ at a future high-energy and high-luminosity
linear electron positron collider, such as the ILC and CLIC to study the sensibility on the anomalous magnetic
and electric dipole moments of the tau-neutrino. For integrated luminosity of 590\hspace{1mm}$fb^{-1}$ and center-of-mass
energy of 3\hspace{1mm}$TeV$, we derive $95 \% \hspace{1mm}C. L.$ limits on the dipole moments:
$\mu_{\nu_\tau}\leq 1.44 \times 10^{-6}\mu_B$ and $d_{\nu_\tau}\leq 2.78 \times 10^{-17}\hspace{1mm} e\hspace{0.5mm} cm$
in the $\gamma^* e^-$ collision mode and of $\mu_{\nu_\tau}\leq 3.4 \times 10^{-7}\mu_B$ and $d_{\nu_\tau}\leq 6.56 \times 10^{-18}\hspace{1mm} e\hspace{0.5mm}cm$ with the $\gamma^*\gamma^*$ collision mode, improving the existing limits.
\end{abstract}


\vspace{5mm}

\maketitle

\section{Introduction}

In the Standard Model (SM) \cite{S.L.Glashow,Weinberg,Salam} extended to contain
right-handed neutrinos, the neutrino magnetic moment induced by radiative corrections
is unobservably small, $\mu_\nu=3eG_F m_{\nu_i}/(8\sqrt{2}\pi^2)\simeq 3.1\times
10^{-19}(m_{\nu_i}/1 \hspace{1mm} eV)\mu_B$, where $\mu_B=e/2m_e$ is the Bohr magneton \cite{Fujikawa,Shrock}.
Current limits on these magnetic moments are several orders of magnitude larger, so that a magnetic moment
close to these limits would indicate a window for probing effects induced by new physics beyond the SM \cite{Fukugita}.
Similarly, a neutrino electric dipole moment will also point to new physics and will be of relevance in astrophysics
and cosmology, as well as terrestrial neutrino experiments \cite{Cisneros}. In the case of the magnetic moment of the
$\nu_e$ the best bound is derived from globular cluster red giants energy loss \cite{Raffelt},

\begin{equation}
\mu_{\nu_e} < 3\times 10^{-12} \mu_B,
\end{equation}

\noindent is far from the SM value. The best current laboratory constraint

\begin{equation}
\mu_{\nu_e} < 2.9\times 10^{-11} \mu_B, \hspace{5mm} 90\%\hspace{0.8mm}C.L.,
\end{equation}

\noindent is obtained in $\bar\nu_e-e^-$ elastic scattering experiment GEMMA \cite{Bed}, which
is an order of magnitude larger than the constraint obtained in astrophysics \cite{Raffelt}.

For the magnetic moment of the muon-neutrino the current best limit has been obtained in
the LSND experiment \cite{Auerbach}

\begin{equation}
\mu_{\nu_\mu} \leq 6.8\times 10^{-10} \mu_B, \hspace{5mm} 90\%\hspace{0.8mm}C.L.
\end{equation}

In the case of the electric dipole moment $d_{\nu_e, \nu_\mu}$ \cite{Aguila} the best limits are:

\begin{equation}
d_{{\nu}_e, {\nu}_\mu} < 2\times 10^{-21} (e cm), \hspace{5mm} 95\%\hspace{0.8mm}C.L.\\
\end{equation}

The most general expression consistent with Lorentz and electromagnetic gauge invariance,
for the tau-neutrino electromagnetic vertex may be parameterized in terms of four form factors:

\begin{equation}
\Gamma^{\alpha}=eF_{1}(q^{2})\gamma^{\alpha}+\frac{ie}{2m_{\nu_\tau}}F_{2}(q^{2})\sigma^{\alpha
\mu}q_{\mu}+eF_3(q^2)\gamma_5\sigma^{\alpha\mu}q_\mu +eF_4(q^2)\gamma_5(\gamma^\mu q^2-q\llap{/}q^\mu),
\end{equation}

\noindent where $e$ is the charge of the electron, $m_{\nu_\tau}$ is the mass of the tau-neutrino, $q^\mu$
is the photon momentum, and $F_{1, 2, 3, 4}(q^2)$ are the electromagnetic form factors of the neutrino,
corresponding to charge radius, magnetic moment (MM), electric dipole moment (EDM) and anapole moment (AM),
respectively, at $q^2=0$ \cite{Escribano,Vogel,Bernabeu1,Bernabeu2,Dvornikov,Giunti,Broggini}. The form factors
corresponding to charge radius and the anapole moment, do not concern us here.

The current best limit on $\mu_{\nu_\tau}$ has been obtained  in the Borexino
experiment which explores solar neutrinos. Searches for the magnetic moment of the
tau-neutrino have also been performed in accelerator experiments. The experiment
E872 (DONUT) is based at $\nu_\tau e^-, \bar\nu_\tau e^-$ elastic scattering.
In the CERN experiment WA-066, a limit on $\mu_{\nu_\tau}$ is obtained on an assumed
flux of tau-neutrinos in the neutrino beam. The L3 collaboration obtain a limit on
the magnetic moment of the tau-neutrino from a sample of $e^+e^-$ annihilation events
at the $Z$ resonance. Experimental limits on the magnetic moment of the tau-neutrino
are shown in Table I.

\begin{table}[!ht]
\caption{Experimental limits on the magnetic moment of the tau-neutrino.}
\begin{center}
\begin{tabular}{c| c| c| c| c}
\hline
Experiment     &  Method          &  Limit                                      & C. L.  &  Reference\\
\hline
\hline

Borexino       &  Solar neutrino  &  $\mu_{\nu_\tau} < 1.9\times 10^{-10}\mu_B$ & $90 \%$  & \cite{Borexino} \\
\hline
E872 (DONUT)   &  Accelerator $\nu_\tau e^-, \bar\nu_\tau e^-$ & $\mu_{\nu_\tau} < 3.9\times 10^{-7}\mu_B$  & $90 \%$  & \cite{DONUT} \\
\hline
CERN-WA-066    &  Accelerator     & $\mu_{\nu_\tau} < 5.4\times 10^{-7}\mu_B$   & $90 \%$ & \cite{A.M.Cooper} \\
\hline
L3             & Accelerator      & $\mu_{\nu_\tau} < 3.3\times 10^{-6}\mu_B$   & $90 \%$  & \cite{L3} \\
\hline
\end{tabular}
\end{center}
\end{table}

Others limits on the magnetic moment of the $\mu_{\nu_\tau}$ are reported in the literature
\cite{Gutierrez9,Gutierrez8,Data2014,Gutierrez7,Gutierrez6,Aydin,Gutierrez5,Perez,Gutierrez4,Gutierrez3,Larios,Keiichi,Aytekin,Gutierrez2,
Hernandez,Maya,Gutierrez1,DELPHI,Escribano,Gould,Grotch}.

In this work we study the sensibility of the anomalous magnetic and electric dipole moments of the
tau-neutrino through the processes $e^+e^- \to e^+\gamma^* e^- \to e^+\tau \bar\nu_\tau \nu_e $
and $e^+e^- \to e^+\gamma^*\gamma^* e^- \to e^+\nu_\tau \bar\nu_\tau e^- $ at a future high-energy
and high-luminosity linear electron positron collider, with a center-of-mass energy in the range
of 500 to 1600 $GeV$, such as the International Linear Collider (ILC) \cite{Abe}, and of 3 $TeV$
to the Compact Linear Collider (CLIC) \cite{Accomando}. Not only can the future $e^{+}e^{-}$ linear collider
be designed to operate in $e^{+}e^{-}$ collision mode, but it can also be operated as a
$e \gamma$ and $\gamma \gamma$ collider. This is achieved by using Compton backscattered photons
in the scattering of intense laser photons on the initial $e^{+}e^{-}$ beams. The other well-known
applications of linear colliders are to study new physics beyond the SM through $e \gamma^{*}$ and
$\gamma^{*}\gamma^{*}$ collisions. A quasi-real $\gamma^{*}$ photon emitted from one of the incoming
$e^{-}$ or $e^{+}$ beams can interact with the other lepton shortly after, and the subprocess
$\gamma^* e^{-}\rightarrow \tau \bar{\nu_{\tau}}\nu_{e}$ can generate. Hence, first, we calculate the main
reaction $e^{+}e^{-} \rightarrow e^{+}\gamma^{*} e^{-} \rightarrow e^{+} \tau \bar{\nu_{\tau}} \nu_{e}$
by integrating the cross section for the subprocess $\gamma^{*} e^{-} \rightarrow \tau \bar{\nu_{\tau}} \nu_{e}$.
Also, $\gamma^{*}$ photons emitted from both $e^{-}$ and $e^{+}$ beams collide with each other, and the
subprocess $\gamma^{*} \gamma^{*} \rightarrow \nu_{\tau} \bar{\nu_{\tau}}$ can be produced. Second, we
find the main reaction $e^{+}e^{-} \rightarrow e^{+}\gamma^{*} \gamma^{*} e^{-} \rightarrow e^{+} \nu_{\tau} \bar{\nu_{\tau}} e^{-}$
by integrating the cross section for the subprocess $\gamma^{*} \gamma^{*} \rightarrow  \nu_{\tau} \bar{\nu_{\tau}} $.
In both cases, the quasi-real photons in $e \gamma^{*}$ and $\gamma^{*}\gamma^{*}$ collisions can be examined
by Equivalent Photon Approximation (EPA) \cite{Budnev,Baur,Piotrzkowski}, that is to say, using the Weizsacker-Williams
approximation. In EPA, photons emitted from incoming leptons which have very low virtuality are scattered
at very small angles from the beam pipe and because the emitted quasi-real photons have a low $Q^{2}$ virtuality,
these are almost real. These processes have been observed experimentally at the LEP, Tevatron and LHC \cite{Abulencia,Aaltonen1,Aaltonen2,Chatrchyan1,Chatrchyan2,Abazov,Chatrchyan3}. In  particular, the most stringent
experimental limit on the anomalous magnetic dipole moment of the tau lepton is obtained through the process
$e^{+}e^{-} \rightarrow e^{+}\gamma^{*} \gamma^{*} e^{-} \rightarrow e^{+} \tau \bar{\tau} e^{-}$ by using
multiperipheral collision at the LEP \cite{DELPHI1}.

In Refs. \cite{Sahin,Sahin1}, the electromagnetic properties of the neutrinos were examined via
the Weizsacker-Williams approximation at the LHC. In Ref. \cite{Sahin} nonstandard couplings
$\nu \bar{\nu} \gamma$ and $\nu \bar{\nu} \gamma \gamma$ were investigated via $\nu \bar{\nu} q$ production
in the process $pp \rightarrow p \gamma^{*} p\rightarrow p \nu \bar{\nu} q X$. In addition, the potential
of $\gamma^{*}\gamma^{*}$ collisions at the LHC was studied via the reaction $pp \rightarrow p\gamma^{*} \gamma^{*} p\rightarrow p \nu \bar{\nu} p$
to probe neutrino-photon coupling by Ref. \cite{Sahin1}.

With these motivations, we study the potential of the processes $e^{+}e^{-} \rightarrow e^{+}\gamma^{*} e^{-} \rightarrow e^{+} \tau \bar{\nu_{\tau}} \nu_{e}$ and $e^{+}e^{-} \rightarrow e^{+}\gamma^{*} \gamma^{*} e^{-} \rightarrow e^{+} \nu_{\tau} \bar{\nu_{\tau}} e^{-}$
and derive limits on the dipole moments $\mu_{\nu_{\tau}}$ and $d_{\nu_{\tau}}$ at
$2\sigma$ and $3\sigma$ level ($90\%$ and $95\%$ C.L.) via Weisacker-Williams approximation,
and at a future high-energy and high-luminosity linear electron positron collider, such as the ILC
and CLIC to study the sensibility on the anomalous magnetic and electric dipole moments of the tau-neutrino.

For this we calculate the main reaction $e^+e^- \to e^+\gamma^* e^- \to e^+\tau \bar\nu_\tau \nu_e $
by integrating the cross section for the subprocess $\gamma^* e^- \to \tau \bar\nu_\tau \nu_e $. The acceptance
cuts will be imposed as $|\eta^\tau|< 2.5$ for pseudorapidity and $p^\tau_T > 20$ $GeV$ for transverse momentum cut
of the final state $\tau$ lepton, respectively. For the second process we calculate the main reaction
$e^+e^- \to e^+\gamma^* \gamma^* e^- \to e^+\nu_\tau \bar\nu_\tau e^- $. Neutrinos in this process are not detected
directly in the central detector. Therefore we do not apply any cuts for the final state particles. The corresponding
Feynman diagrams for the main reactions as well as for the sub-processes which give the most important contribution to
the total cross-section are shown in Figs. 1-4.

To illustrate our results for both processes we show the dependence of the total cross-section as a function
of anomalous couplings $F_2$ and $F_3$ for three different values of the center-of-mass energies $0.5, 1.5$
and $3$ $TeV$, respectively. The variation of the cross-section as a function of $F_2$ and $F_3$ for different
values of $Q^2$ (Weizsacker-Williams photon virtuality) and center-of-mass energy of $0.5, 1.5$ and $3$ $TeV$
is evaluated. We also include a contours plot for the upper bounds of the anomalous couplings $\mu_{\nu_\tau}$
and $d_{\nu_\tau}$ with $95 \%$ C.L. at the $\sqrt{s}=0.5, 1.5, 3$ $TeV$ with corresponding maximum luminosities
for both processes. The sensitivity limits on the magnetic moment $\mu_{\nu_\tau}$ and the electric dipole moment
$d_{\nu_\tau}$ of the tau-neutrino for different values of photon virtuality, center-of-mass energy and luminosity
are also calculated.

This paper is organized as follows. In Section II, we study the dipole moments of the tau-neutrino through the
processes $e^+e^- \to e^+\gamma^* e^- \to e^+\tau \bar\nu_\tau \nu_e $ in the $\gamma^* e^-$ collision mode and
$e^+e^- \to e^+\gamma^* \gamma^* e^- \to e^+\nu_\tau \bar\nu_\tau e^- $ through the $\gamma^*\gamma^*$
collision mode. Finally, we present our results and conclusions in Section III.

\vspace{5mm}

\section{Cross-section of $e^+e^- \to e^+\gamma^* e^- \to e^+\tau \bar\nu_\tau \nu_e $ and
$e^+e^- \to e^+\gamma^* \gamma^* e^- \to e^+\nu_\tau \bar\nu_\tau e^- $}

\vspace{3mm}

In this section we present numerical results of the cross-section for both processes  $e^+e^- \to e^+\gamma^* e^- \to e^+\tau \bar\nu_\tau \nu_e $
and $e^+e^- \to e^+\gamma^* \gamma^* e^- \to e^+\nu_\tau \bar\nu_\tau e^- $ as a function of the electromagnetic
form factors of the neutrino $F_2$ and $F_3$. In addition, to see the sensitivity of the magnetic moment
$\mu_{\nu_\tau}$ and the electric dipole moment $d_{\nu_\tau}$ to new physics, we plot $\mu_{\nu_\tau} (d_{\nu_\tau} )$
versus ${\cal L}$. We carry out the calculations using the framework
of the minimally extended standard model at next generation linear $\gamma^* e^-$ and $\gamma^*\gamma^*$
collisions: ILC and CLIC.

We use the CompHEP \cite{Pukhov} packages for calculations of the matrix elements and
cross-sections. These packages provide automatic computation of the cross-sections and
distributions in the SM as well as their extensions at tree-level. We consider the
high energy stage of possible future linear $\gamma^* e^-$ and $\gamma^*\gamma^*$
collisions with $\sqrt{s}=0.5, 1.5$ and 3 $TeV$ and design luminosity 230, 320 and 590
$fb^{-1}$ according to the data reported by the ILC and CLIC \cite{Abe,Accomando}. In
addition, we consider the acceptance cuts of $|\eta^\tau|< 2.5$ for pseudorapidity and
$p^\tau_T > 20$ $GeV$ for transverse momentum cut of the final state $\tau$ lepton, respectively.

\subsection{ Magnetic moment and electric dipole moment via $e^+e^- \to e^+\gamma^* e^- \to e^+\tau \bar\nu_\tau \nu_e$}

The corresponding Feynman diagrams for the main reaction $e^+e^- \to e^+\gamma^* e^- \to e^+\tau \bar\nu_\tau \nu_e$,
as well as for the subprocess $\gamma^* e^- \to \tau \bar\nu_\tau \nu_e $ which give the most important contribution to
the total cross-section are shown in Figs. 1-2. From Fig. 2, the Feynman diagrams (1)-(3) correspond to the contribution of
the standard model, while diagram (4) corresponds to the anomalous contribution, that is to say, for the $\gamma^*e^-$
collisions there are SM background at the tree-level so the total cross-section is proportional to
$\sigma_{Tot}=\sigma_{SM}+\sigma_{Int}(F_2, F_3)+\sigma_{Anom}(F^2_2, F^2_3, F_2F_3)$, respectively.

To illustrate our results we show the dependence of the cross-section on the anomalous
couplings $F_2$ and $F_3$ for $e^+e^- \to e^+\gamma^* e^- \to e^+\tau \bar\nu_\tau \nu_e $
in Fig. 5 for three different center-of-mass energies $\sqrt{s}=0.5, 1.5, 3$\hspace{1mm} $TeV$
and $Q^2 = 2, 16, 64$\hspace{1mm} $GeV^2$ \cite{Dvornikov}, respectively. The cross-section is sensitive to the
value of the center-of-mass energies, as well as to $Q^2$. The sensitivity to $e^+\tau \bar\nu_\tau \nu_e $
increases with the collider energy, as well as with $Q^2$ reaching a maximum at the end of the
range considered: $F_{2, 3}=\pm 0.001$. In Fig. 6, we show again the total cross-section, but now for
different values ​​of $Q^2 = 2, 16, 64$\hspace{1mm} $GeV^2$ \cite{Dvornikov} and center-of-mass energies  of
$\sqrt{s}=0.5, 1.5, 3$\hspace{1mm} $TeV$. We observed that the variation of the cross-section for
$e^+\tau \bar\nu_\tau \nu_e $ as a function of the anomalous couplings $F_2$ and $F_3$ it is clear
for all case.

In Figures 7 and 8 we present the dependence of the sensitivity limits of the magnetic
moment $\mu_{\nu_\tau}$ and the electric dipole moment $d_{\nu_\tau}$ with respect to the collider
luminosity ${\cal L}$ for three different values of the Weizsacker-Williams photon virtuality
$Q^2 = 2, 16, 64$\hspace{1mm} $GeV^2$ and center-of-mass energies of $\sqrt{s}=0.5, 1.5, 3$\hspace{1mm} $TeV$.
In these figures, we observe one variation of $\mu_{\nu_\tau} (d_{\nu_\tau} )$ in all the interval of ${\cal L}$,
and it is almost independent of the value of $Q^2$.

As an indicator of the order of magnitude, in Tables II-III we present the bounds obtained on the
$\mu_{\nu_\tau}$ magnetic moment and $d_{\nu_\tau}$ electric dipole moment for $Q^2 = 2, 64$\hspace{1mm}$GeV^2$,
$\sqrt{s}= 0.5, 1.5, 3$\hspace{1mm}$TeV$ and ${\cal L} = 230, 320, 590$\hspace{0.8mm}$fb^{-1}$ at $2\sigma$
and $3\sigma$ $C. L.$, respectively. We observed that the results obtained in Tables II and III are competitive with those
reported in the literature \cite{DONUT,A.M.Cooper,L3}. For the electric dipole moment our limits compare favorably
with those reported by K. Akama, {\it et al.} \cite{Keiichi} $|d_{\nu_\tau}| < O(2\times 10^{-17}\hspace{0.8mm}ecm)$
and R. Escribano, {\it et al.} \cite{Escribano} $|d_{\nu_\tau}| \leq 5.2\times 10^{-17}\hspace{0.8mm}ecm$, $95\%\hspace{1mm} C. L.$

In Fig. 9 we used three center-of-mass energies $\sqrt{s}=0.5, 1.5, 3\hspace{1mm}TeV $ planned for the ILC and CLIC accelerators
in order to get contours limits in the plane $\mu_{\nu_\tau}-d_{\nu_\tau}$ for
$e^+e^- \to e^+\gamma^* e^- \to e^+\tau \bar\nu_\tau \nu_e $ and the planned luminosities of ${\cal L}=230, 320, 590\hspace{1mm} fb^{-1}$
and Weizsacker-Williams photon virtuality $Q^2 = 2$\hspace{1mm}$GeV^2$. For the $\gamma^* e^-$ collision, we perform $\chi^2$
analysis at $95 \% \hspace{1mm} C. L.$ since the number of SM events is greater than 10.

\begin{table}[!ht]
\caption{Bounds on the $\mu_{\nu_\tau}$ magnetic moment and $d_{\nu_\tau}$ electric dipole moment for the process $e^+e^- \to e^+\gamma^* e^- \to e^+\tau \bar\nu_\tau \nu_e $ for $Q^2 = 2$\hspace{1mm}$GeV^2$,  $\sqrt{s}=0.5, 1.5, 3\hspace{0.8mm}TeV$ and ${\cal L}=230, 320, 590\hspace{0.8mm}fb^{-1}$ at $2\sigma$ and $3\sigma$ C. L.}
\begin{center}
 \begin{tabular}{ccc}
\hline\hline
\multicolumn{3}{c}{$\sqrt{s}=0.5, \hspace{0.8mm} 1.5,\hspace{0.8mm} 3\hspace{0.8mm}TeV$, \hspace{3mm} ${\cal L}=230, 320, 590\hspace{0.8mm}fb^{-1}$}\\
 \hline
 \cline{1-3} C. L.          & $|\mu_{\nu_\tau}(\mu_B)|$        & $|d_{\nu_\tau}(e cm)|$ \\
\hline
$2\sigma$                   & (8.73, 3.35, 1.60)$\times 10^{-6}$                   & \hspace{2mm} (16.8, 6.46, 3.08)$\times 10^{-17}$ \\
$3\sigma$                   & (9.30, 3.30, 1.53)$\times 10^{-6}$                   & \hspace{2mm} (17.9, 6.36, 2.95)$\times 10^{-17}$ \\
\hline\hline
\end{tabular}
\end{center}
\end{table}

\begin{table}[!ht]
\caption{Bounds on the $\mu_{\nu_\tau}$ magnetic moment and $d_{\nu_\tau}$ electric dipole moment for the process $e^+e^- \to e^+\gamma^* e^- \to e^+\tau \bar\nu_\tau \nu_e $ for $Q^2 = 64$\hspace{1mm}$GeV^2$,  $\sqrt{s}=0.5, 1.5, 3\hspace{0.8mm}TeV$ and ${\cal L}=230, 320, 590\hspace{0.8mm}fb^{-1}$ at $2\sigma$ and $3\sigma$ C. L.}
\begin{center}
 \begin{tabular}{ccc}
\hline\hline
\multicolumn{3}{c}{$\sqrt{s}=0.5, \hspace{0.8mm} 1.5,\hspace{0.8mm} 3\hspace{0.8mm}TeV$, \hspace{3mm} ${\cal L}=230, 320, 590\hspace{0.8mm}fb^{-1}$}\\
 \hline
 \cline{1-3} C. L.          & $|\mu_{\nu_\tau}(\mu_B)|$        & $|d_{\nu_\tau}(e cm)|$ \\
\hline
$2\sigma$                   & (8.22, 2.88, 1.32)$\times 10^{-6}$                   & \hspace{2mm} (15.8, 5.56, 2.54)$\times 10^{-17}$ \\
$3\sigma$                   & (8.97, 3.14, 1.44)$\times 10^{-6}$                   & \hspace{2mm} (17.3, 6.06, 2.78)$\times 10^{-17}$ \\
\hline\hline
\end{tabular}
\end{center}
\end{table}

\subsection{ Magnetic moment and electric dipole moment via $e^+e^- \to e^+\gamma^* \gamma^* e^- \to e^+\nu_\tau \bar\nu_\tau e^- $}

In this subsection we study the dipole moments of the tau-neutrino via the process
$e^+e^- \to e^+\gamma^* \gamma^* e^- \to e^+\nu_\tau \bar\nu_\tau e^- $ for energies expected
at the ILC and CLIC \cite{Abe,Accomando}. The corresponding Feynman diagrams for the subprocess
$\gamma^* \gamma^* \to \nu_\tau \bar\nu_\tau$ which give the most important contribution to the total
cross-section are shown in Figs. 3 and 4. In this case, the total cross-section of the subprocess depends
only on the diagrams (1) and (2) with anomalous couplings, and there is no contribution at tree level of the
standard model, that is to say $\sigma_{Tot}=\sigma(F^4_2, F^4_3, F^3_2F_3, F^2_2F^2_3, F_2F^3_3)$.

For the study of the subprocess $\gamma^* \gamma^* \to \nu_\tau \bar\nu_\tau$ in Fig. 10, we show the
total cross-section as a function of the electromagnetic form factors of the neutrino $F_2$ and $F_3$ for
three different center-of-mass energies $\sqrt{s} = 0.5, 1.5, 3$\hspace{1mm}$TeV$ and $Q^2 = 2, 16, 64$\hspace{1mm}$GeV^2$ \cite{Dvornikov},
respectively. We can see from this figure that the total cross-section changes strongly with the variation
of the $\sqrt{s}$ and $Q^2$ values.

As in subsection A, we show the $F_2$ and $F_3$ dependence of the total cross-section for
$e^+e^- \to e^+\gamma^* \gamma^* e^- \to e^+\nu_\tau \bar\nu_\tau e^- $ in Fig. 11. From this
figure we observed a significant dependence of the cross-section with respect to $F_2$
and $F_3$, and different values of center-of-mass energy $\sqrt{s}$ and $Q^2$. In Figures 12
and  13 we present the dependence of the sensitivity limits of the magnetic moment $\mu_{\nu_\tau}$
and the electric dipole moment $d_{\nu_\tau}$ with respect to the collider luminosity ${\cal L}$ for
three different values of $Q^2 = 2, 16, 64$\hspace{1mm} $GeV^2$ \cite{Dvornikov} and center-of-mass
energies of $\sqrt{s}=0.5, 1.5, 3$\hspace{1mm} $TeV$.

In Tables IV and V we present the bounds obtained on the $\mu_{\nu_\tau}$ magnetic
moment and $d_{\nu_\tau}$ electric dipole moment for $\sqrt{s}= 0.5, 1.5, 3$\hspace{1mm}$TeV$,
$Q^2 = 2, 64$\hspace{1mm} $GeV^2$ and ${\cal L} = 230, 320, 590$\hspace{1mm}$fb^{-1}$ at $2\sigma$
and $3\sigma$. We observed that the results obtained in Tables IV-V improve the bounds
reported in the literature \cite{DONUT,A.M.Cooper,L3}.

In the case of the electric dipole moment the $90, 95\%$ C. L. sensitivity limits at 0.5, 1.5 and
3\hspace{0.8mm}$TeV$  center of mass energies and integrated luminosities of 230, 320 and
$590\hspace{0.8mm}fb^{-1}$, respectively can provide proof of these bounds of order $10^{-18}$,
that is to say, they are improved by one order of magnitude than those reported in the literature:
$|d_{\nu_\tau}| < O(2\times 10^{-17}\hspace{0.8mm}ecm)$ \cite{Keiichi} and
$|d_{\nu_\tau}| \leq 5.2\times 10^{-17}\hspace{0.8mm}ecm$, $95\%$ C. L. \cite{Escribano}.

Finally, in Fig. 14 we summarize the respective limit contours for the dipole moments in the
$\mu_{\nu_\tau}-d_{\nu_\tau}$ plane for $e^+e^- \to e^+\gamma^* \gamma^* e^- \to e^+ \nu_\tau \bar \nu_{\tau} e^-$.
Starting from the top, the curves are for $\sqrt{s}=0.5$\hspace{1mm}$TeV$ and ${\cal L}=230$ $fb^{-1}$;
$\sqrt{s}=1.5$\hspace{1mm}$TeV$ and ${\cal L}=320$ $fb^{-1}$; $\sqrt{s}=3$\hspace{1mm}$TeV$ and
${\cal L}=590$ $fb^{-1}$, respectively. We have used $Q^2=2$\hspace{1mm}$GeV^2$. In this case for the $\gamma^* \gamma^*$
collision, we perform Poisson analysis at $95 \% \hspace{1mm} C. L.$ since the number of SM events is smaller than 10.

\begin{table}[!ht]
\caption{Bounds on the $\mu_{\nu_\tau}$ magnetic moment and $d_{\nu_\tau}$ electric dipole moment for the process $e^+e^- \to e^+\gamma^* \gamma^* e^- \to e^+\nu_\tau \bar\nu_\tau e^- $ for $Q^2 = 2$\hspace{1mm}$GeV^2$,  $\sqrt{s}=0.5, 1.5, 3\hspace{0.8mm}TeV$ and ${\cal L}=230, 320, 590\hspace{0.8mm}fb^{-1}$ at $2\sigma$ and $3\sigma$ C. L.}
\begin{center}
 \begin{tabular}{ccc}
\hline\hline
\multicolumn{3}{c}{$\sqrt{s}=0.5, \hspace{0.8mm} 1.5,\hspace{0.8mm} 3\hspace{0.8mm}TeV$, \hspace{3mm} ${\cal L}=230, 320, 590\hspace{0.8mm}fb^{-1}$}\\
 \hline
 \cline{1-3} C. L.          & $|\mu_{\nu_\tau}(\mu_B)|$        & $|d_{\nu_\tau}(e cm)|$ \\
\hline
$2\sigma$                   & (10.90, 5.70, 3.50)$\times 10^{-7}$                   & \hspace{2mm} (2.10, 1.09)$\times 10^{-17}$, 6.75$\times 10^{-18}$ \\
$3\sigma$                   & (11.60, 6.10, 3.70)$\times 10^{-7}$                   & \hspace{2mm} (2.24, 1.18)$\times 10^{-17}$, 7.14$\times 10^{-18}$ \\
\hline\hline
\end{tabular}
\end{center}
\end{table}

\begin{table}[!ht]
\caption{Bounds on the $\mu_{\nu_\tau}$ magnetic moment and $d_{\nu_\tau}$ electric dipole moment for the process $e^+e^- \to e^+\gamma^* \gamma^* e^- \to e^+\nu_\tau \bar\nu_\tau e^- $ for $Q^2 = 64$\hspace{1mm}$GeV^2$,  $\sqrt{s}=0.5, 1.5, 3\hspace{0.8mm}TeV$ and ${\cal L}=230, 320, 590\hspace{0.8mm}fb^{-1}$ at $2\sigma$ and $3\sigma$ C. L.}
\begin{center}
 \begin{tabular}{ccc}
\hline\hline
\multicolumn{3}{c}{$\sqrt{s}=0.5, \hspace{0.8mm} 1.5,\hspace{0.8mm} 3\hspace{0.8mm}TeV$, \hspace{3mm} ${\cal L}=230, 320, 590\hspace{0.8mm}fb^{-1}$}\\
 \hline
 \cline{1-3} C. L.          & $|\mu_{\nu_\tau}(\mu_B)|$        & $|d_{\nu_\tau}(e cm)|$ \\
\hline
$2\sigma$                   & (9.90, 5.20, 3.10)$\times 10^{-7}$                   & \hspace{2mm} (1.91, 1.00)$\times 10^{-17}$, 5.98$\times 10^{-18}$ \\
$3\sigma$                   & (10.60, 5.54, 3.40)$\times 10^{-7}$                   & \hspace{2mm} (2.04, 1.07)$\times 10^{-17}$, 6.56$\times 10^{-18}$ \\
\hline\hline
\end{tabular}
\end{center}
\end{table}

\vspace{5cm}

\section{Conclusions}

\vspace{3mm}

Even though $\gamma e^-$ and $\gamma \gamma$ processes require new equipment, $\gamma^{*} e^-$ and $\gamma^{*} \gamma^{*}$
are realized spontaneously at linear colliders without any equipment. These processes will allow the next
generation linear collider to operate in three different modes, $e^+e^-$, $\gamma^* e^-$ and $\gamma^*\gamma^*$,
opening up the opportunity for a wider search for new physics. Therefore, the $\gamma^* e^-$ and $\gamma^*\gamma^*$
linear collisions represents an excellent opportunity to study the sensibility on the anomalous magnetic moment
and electric dipole moment of the tau-neutrino.

We have done an analysis of the total cross-section of the  processes $e^+e^- \to e^+\gamma^* e^- \to e^+\tau \bar\nu_\tau \nu_e $
and $e^+e^- \to e^+\gamma^* \gamma^* e^- \to e^+\nu_\tau \bar\nu_\tau e^-$ as a function of the anomalous coupling $F_2$ and $F_3$.
The analysis is shown in Figs. 5, 6, 10 and 11 for different center-of-mass energies and several values of the Weizsacker-Williams
photon virtuality. In all cases, the cross-section shows a strong dependence on the anomalous couplings $F_2$ and $F_3$.

The correlation between the luminosity ${\cal L}$ of the collider and the anomalous magnetic moment $\mu_{\nu_\tau}$
and the electric dipole moment $d_{\nu_\tau}$ is presented in Figs. 7 and 8. In both cases, we see that there is a strong
correlation between ${\cal L}$ and the dipole moments, the same is also observed in Figs. 12 and 13 as well as in Tables II-V.

We also include contours plots for the dipole moments at the $95 \%\hspace{1mm}C. L.$ in the $\mu_{\nu_\tau}-d_{\nu_\tau}$
plane for $Q^2= 2\hspace{1mm}GeV^2$ and $\sqrt{s}=0.5, 1.5, 3\hspace{1mm}TeV$ in Figures 9 and 14. The contours are obtained
from Tables II-V.

It is worth mentioning that our bounds obtained in Tables II-V on the anomalous magnetic moment for the processes
$e^+e^- \to e^+\gamma^* e^- \to e^+\tau \bar\nu_\tau \nu_e $ and $e^+e^- \to e^+\gamma^* \gamma^* e^- \to e^+\nu_\tau \bar\nu_\tau e^-$
for $Q^2 = 2, 64$\hspace{1mm}$GeV^2$, $\sqrt{s}=0.5, 1.5, 3\hspace{0.8mm}TeV$ and ${\cal L}=230, 320, 590\hspace{0.8mm}fb^{-1}$ at $2\sigma$
and $3\sigma$ C. L. compare favorably with the bounds obtained in Table I by DONUT \cite{DONUT}, WA66 \cite{A.M.Cooper}
and L3 Collaboration \cite{L3}, as well as those reported by K. Akama, {\it et al.} \cite{Keiichi} $\mu_{\nu_\tau}< O(1.1\times 10^{-6}\hspace{1mm}\mu_B)$
and R. Escribano {\it et al.} \cite{Escribano} $\mu_{\nu_\tau}\leq 2.7\times 10^{-6}\hspace{1mm}\mu_B$ $95\%\hspace{1mm}C. L.$
While in the case of the electric dipole moment our results obtained in Table II-V are improved by one order of magnitude than those
reported in the literature $|d_{\nu_\tau}| < O(2\times 10^{-17}\hspace{1mm}e cm)$ \cite{Keiichi} and $|d_{\nu_\tau}| \leq 5.2\times 10^{-17}\hspace{1mm}e cm$, $95\%\hspace{1mm}C. L.$ \cite{Escribano}.

In conclusion, we have found that the processes $e^+e^- \to e^+\gamma^* e^- \to e^+\tau \bar\nu_\tau \nu_e $
and $e^+e^- \to e^+\gamma^* \gamma^* e^- \to e^+\nu_\tau \bar\nu_\tau e^-$ in the $\gamma^* e^-$ and $\gamma^*\gamma^*$
collision modes at the high energies and luminosities expected at the ILC and CLIC colliders can be used to probe for bounds
on the magnetic moment $\mu_{\nu_\tau}$ and electric dipole moment $d_{\nu_\tau}$ of the tau-neutrino. In particular,
we can appreciate that for integrated luminosities of 590\hspace{1mm}$fb^{-1}$ and center-of-mass energies
of 3\hspace{1mm}$TeV$, we derive $95 \% \hspace{1mm}C. L.$ limits on the dipole moments:
$\mu_{\nu_\tau}\leq 1.44 \times 10^{-6}\mu_B$ and $d_{\nu_\tau}\leq 2.78 \times 10^{-17}\hspace{1mm} e\hspace{0.5mm} cm$
for the process $e^+e^- \to e^+\gamma^* e^- \to e^+\tau \bar\nu_\tau \nu_e $ and of
$\mu_{\nu_\tau}\leq 3.4 \times 10^{-7}\mu_B$ and $d_{\nu_\tau}\leq 6.56 \times 10^{-18}\hspace{1mm} e\hspace{0.5mm}cm$
for $e^+e^- \to e^+\gamma^* \gamma^* e^- \to e^+\nu_\tau \bar\nu_\tau e^- $, better than those reported in the literature.

\vspace{8mm}

\begin{center}
{\bf Acknowledgements}
\end{center}

A. G. R. acknowledges support from CONACyT, SNI, PROMEP and PIFI (M\'exico).

\newpage

\begin{figure}[t]
\centerline{\scalebox{0.69}{\includegraphics{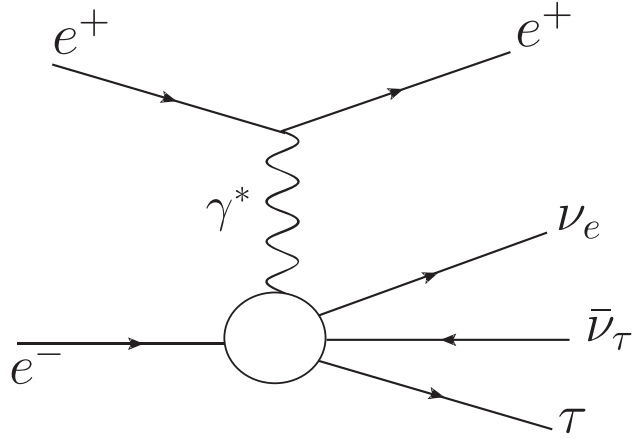}}}
\caption{ \label{fig:gamma} Schematic diagram for the process
$e^+e^- \to e^+\gamma^* e^- \to e^+\tau \bar\nu_\tau \nu_e $.}
\end{figure}

\begin{figure}[t]
\centerline{\scalebox{0.85}{\includegraphics{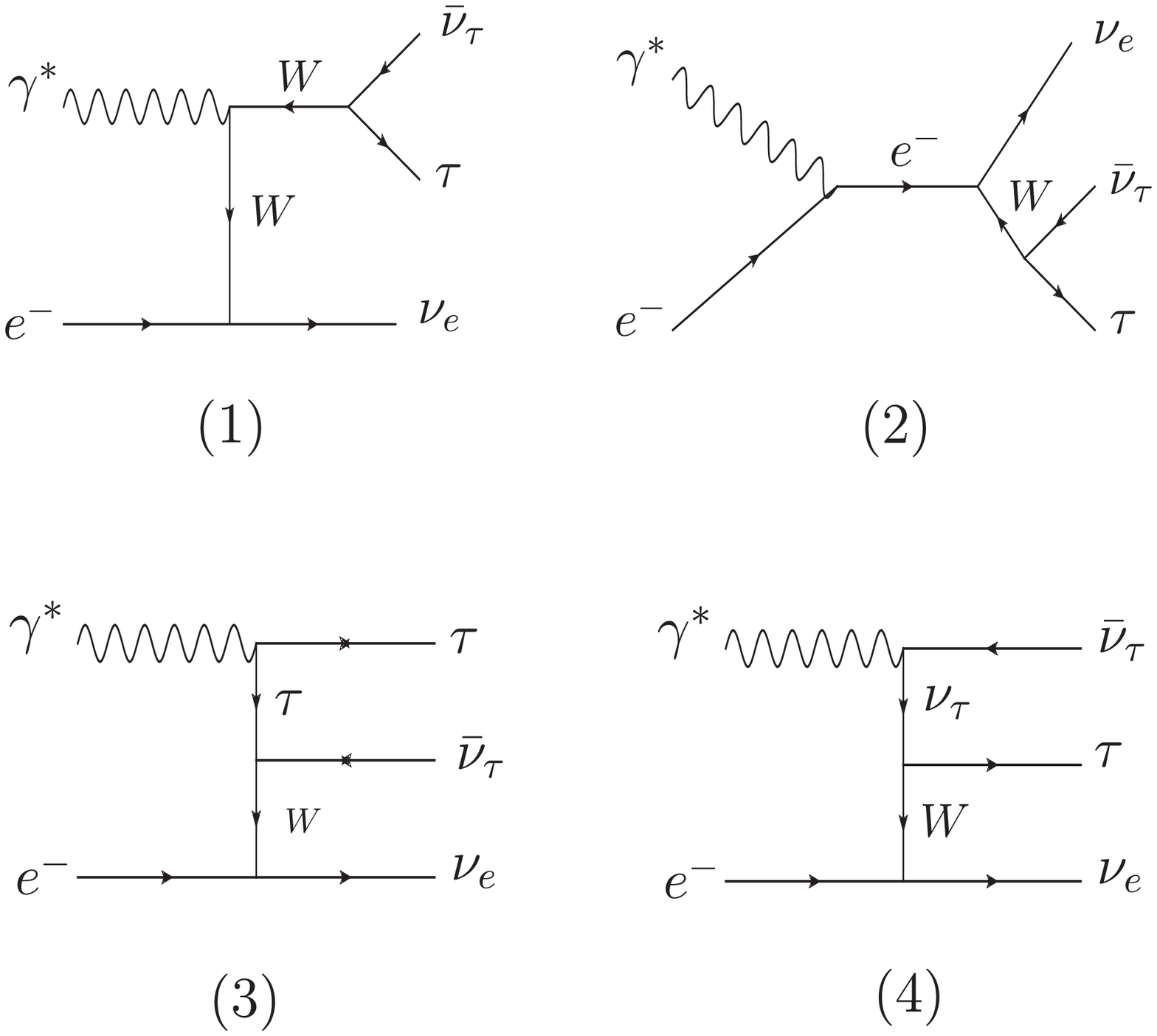}}}
\caption{ \label{fig:gamma} The Feynman diagrams contributing to the subprocess
$\gamma^* e^- \to \tau \bar\nu_\tau \nu_e $.}
\end{figure}

\begin{figure}[t]
\centerline{\scalebox{0.69}{\includegraphics{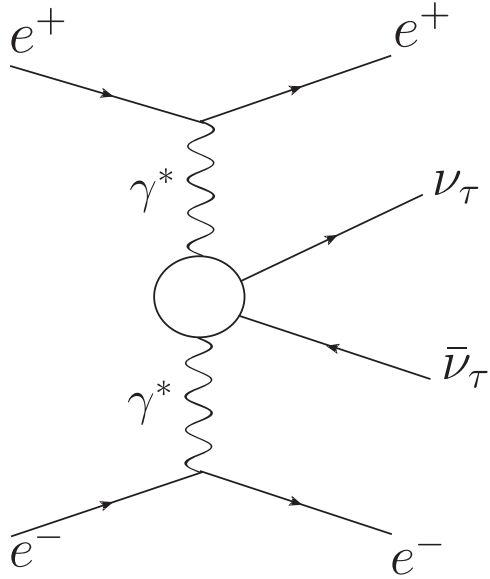}}}
\caption{ \label{fig:gamma1} Schematic diagram for the process
$e^+e^- \to e^+\gamma^* \gamma^* e^- \to e^+\nu_\tau \bar\nu_\tau e^- $.}
\end{figure}

\begin{figure}[t]
\centerline{\scalebox{0.8}{\includegraphics{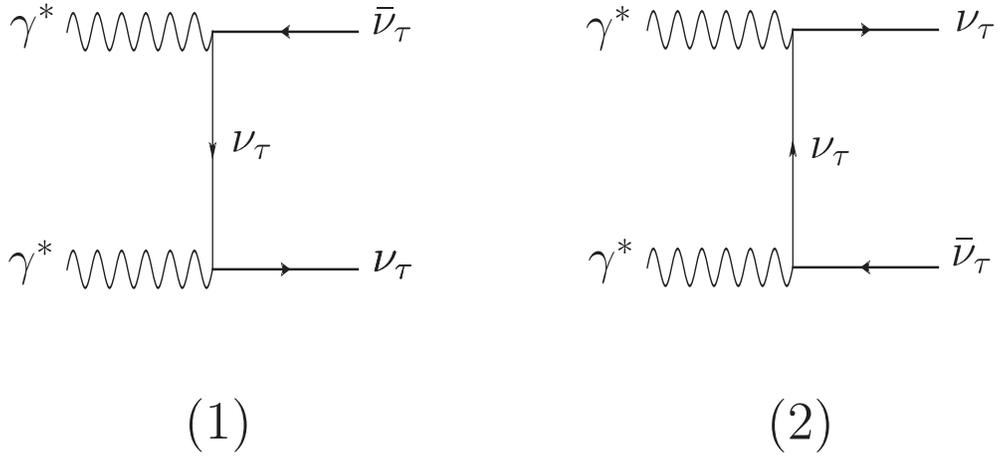}}}
\caption{ \label{fig:gamma2} The Feynman diagrams contributing to the subprocess
$\gamma^*\gamma^* \to \nu_\tau \bar\nu_\tau $.}
\end{figure}

\begin{figure}[t]
\centerline{\scalebox{0.8}{\includegraphics{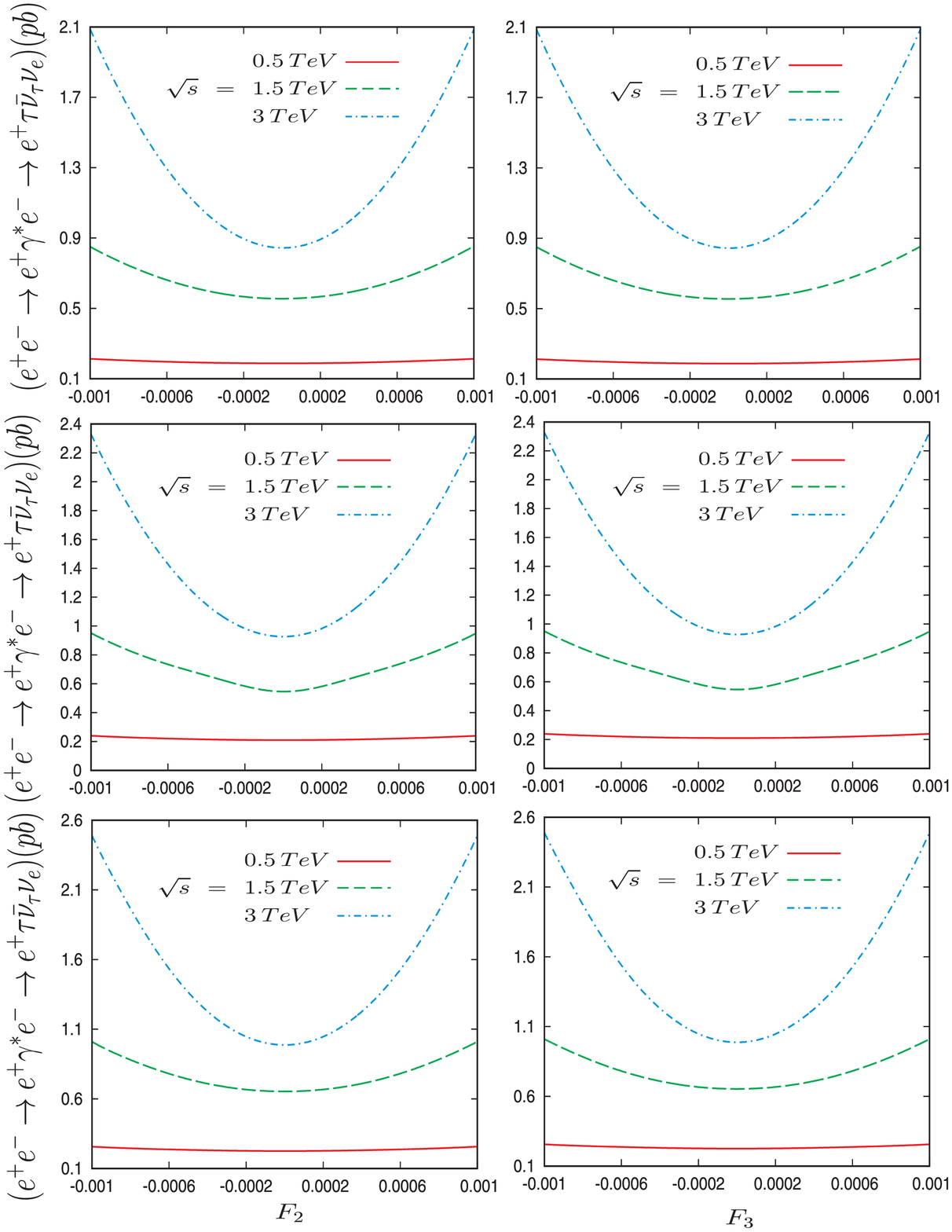}}}
\caption{ \label{fig:gamma3} The integrated total cross-section of the process
$e^+e^- \to e^+\gamma^* e^- \to e^+\tau \bar\nu_\tau \nu_e $ as a function of the
anomalous couplings $F_2$ and $F_3$ for three different center-of-mass energies
$\sqrt{s}=0.5, 1.5, 3$\hspace{1mm}$TeV$ and $Q^2=2, 16, 64$\hspace{1mm}$GeV^2$,
respectively.}
\end{figure}

\begin{figure}[t]
\centerline{\scalebox{0.77}{\includegraphics{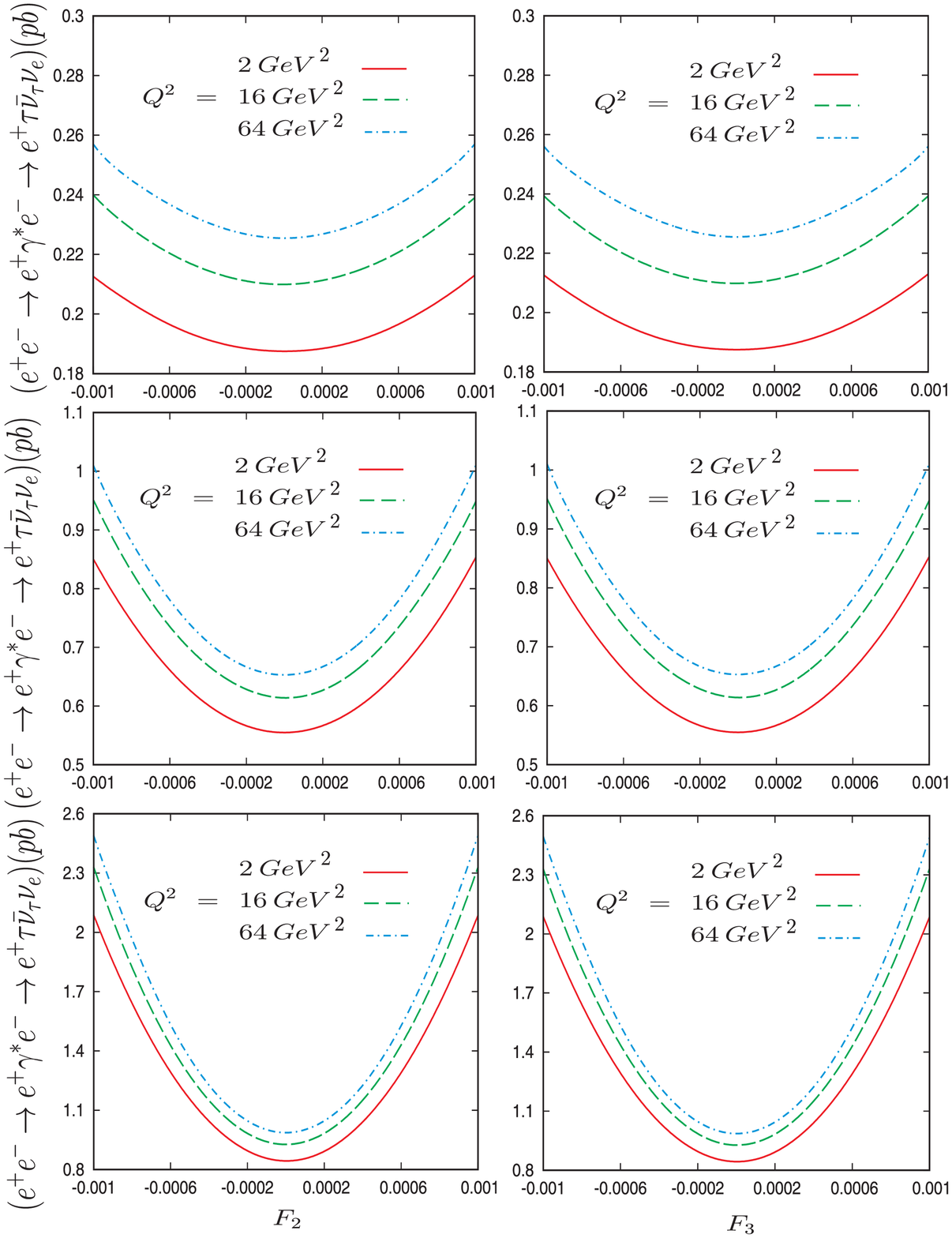}}}
\caption{ \label{fig:gamma4} The total cross section of the process
$e^+e^- \to e^+\gamma^* e^- \to e^+\tau \bar\nu_\tau \nu_e $ as a function of
the anomalous couplings $F_2$ and $F_3$ for three different values of
$Q^2=2, 16, 64$\hspace{1mm}$GeV^2$ and center-of-mass energies
$\sqrt{s}=0.5, 1.5, 3$\hspace{1mm}$TeV$, respectively.}
\end{figure}

\begin{figure}[t]
\centerline{\scalebox{0.77}{\includegraphics{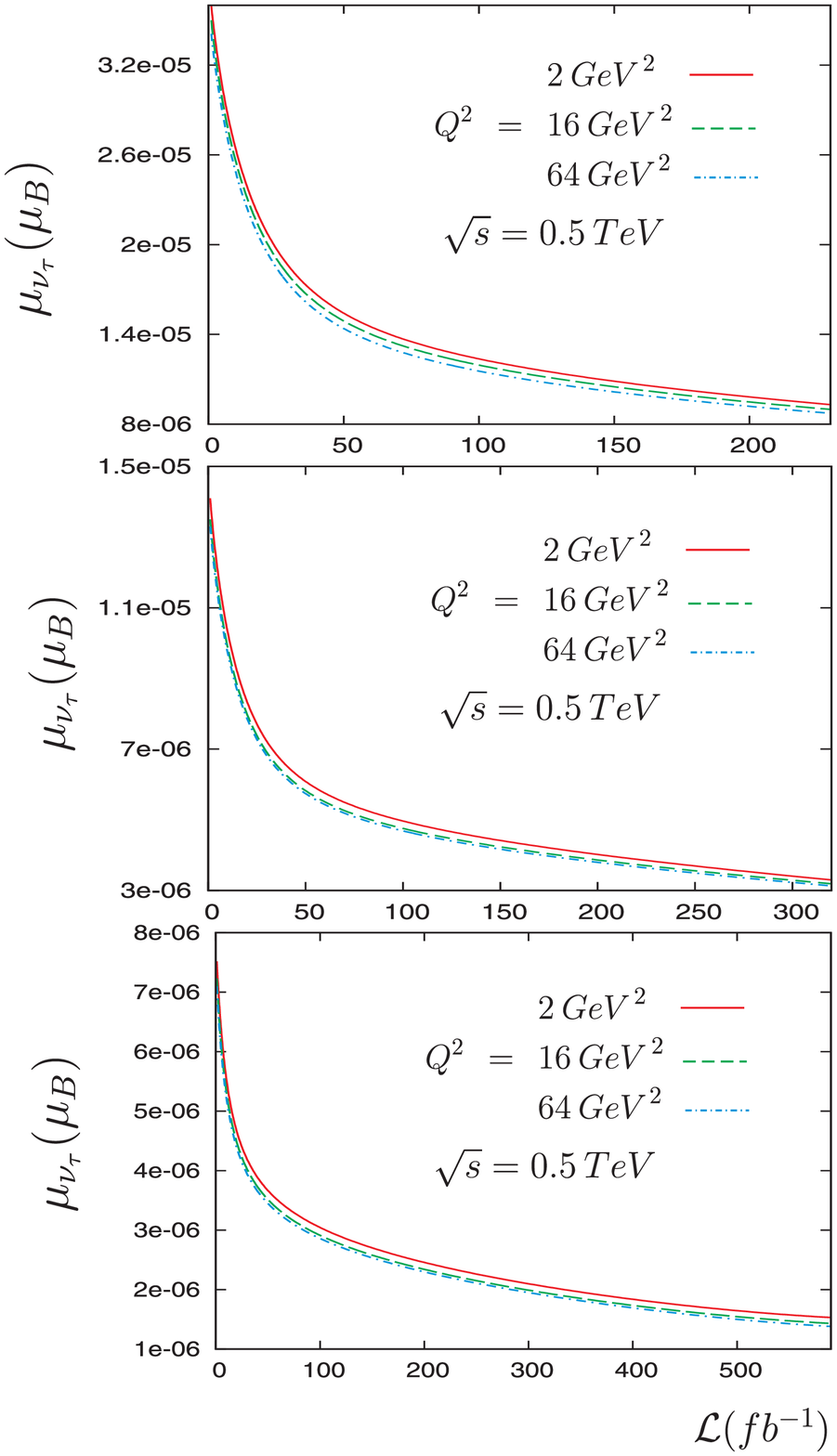}}}
\caption{ \label{fig:gamma15} Dependence of the sensitivity limits at $95\% \hspace{1mm}C. L.$ for the
anomalous magnetic moment for three different values of $Q^2=2, 16, 64$\hspace{1mm}$GeV^2$
and center-of-mass energies $\sqrt{s}=0.5, 1.5, 3$\hspace{1mm}$TeV$ in the subprocess
$\gamma^* e^- \to \tau \bar\nu_\tau \nu_e$.}
\end{figure}

\begin{figure}[t]
\centerline{\scalebox{0.77}{\includegraphics{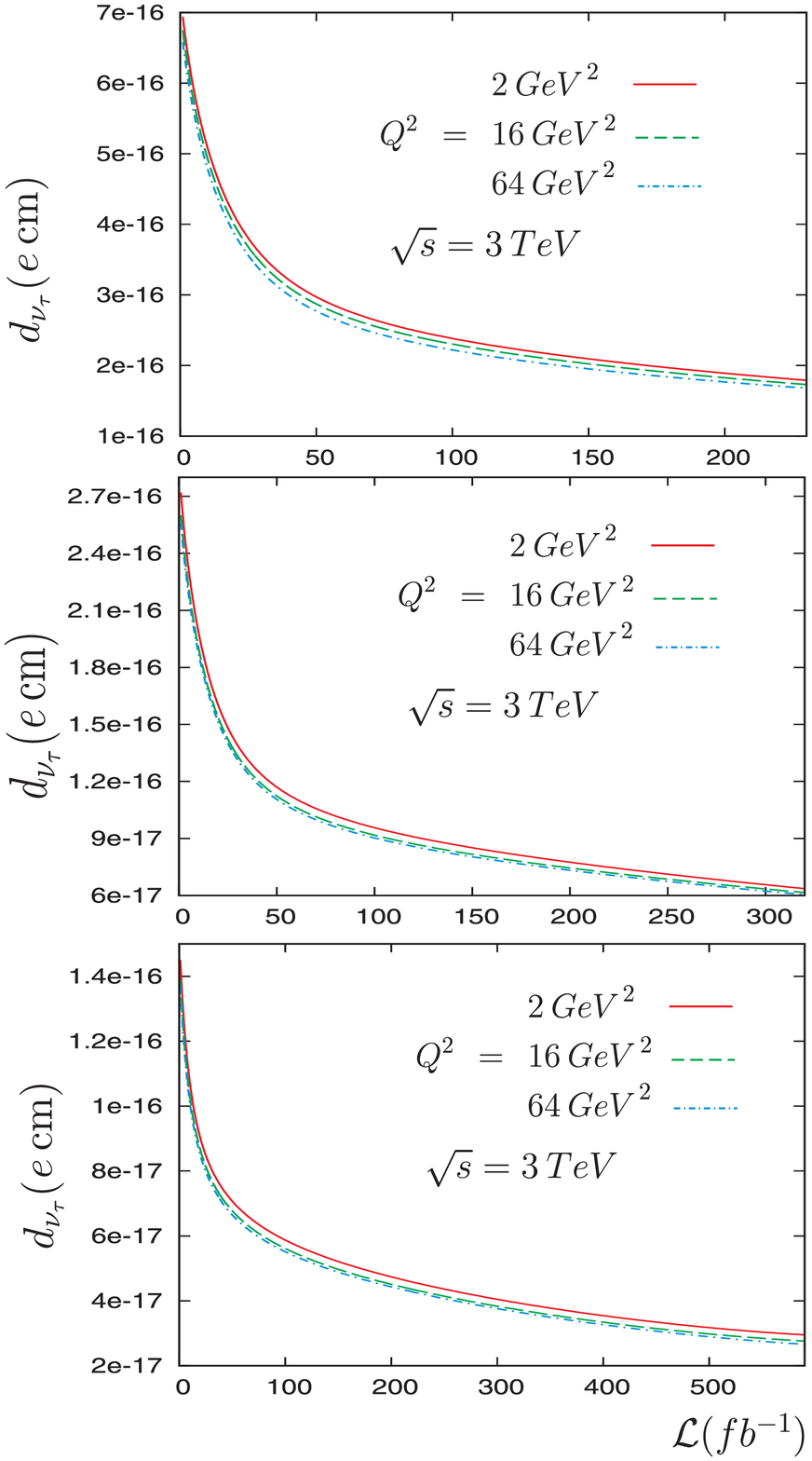}}}
\caption{\label{fig:gamma16} The same as Fig. 7 but for the electric dipole moment.}
\end{figure}

\begin{figure}[t]
\centerline{\scalebox{0.65}{\includegraphics{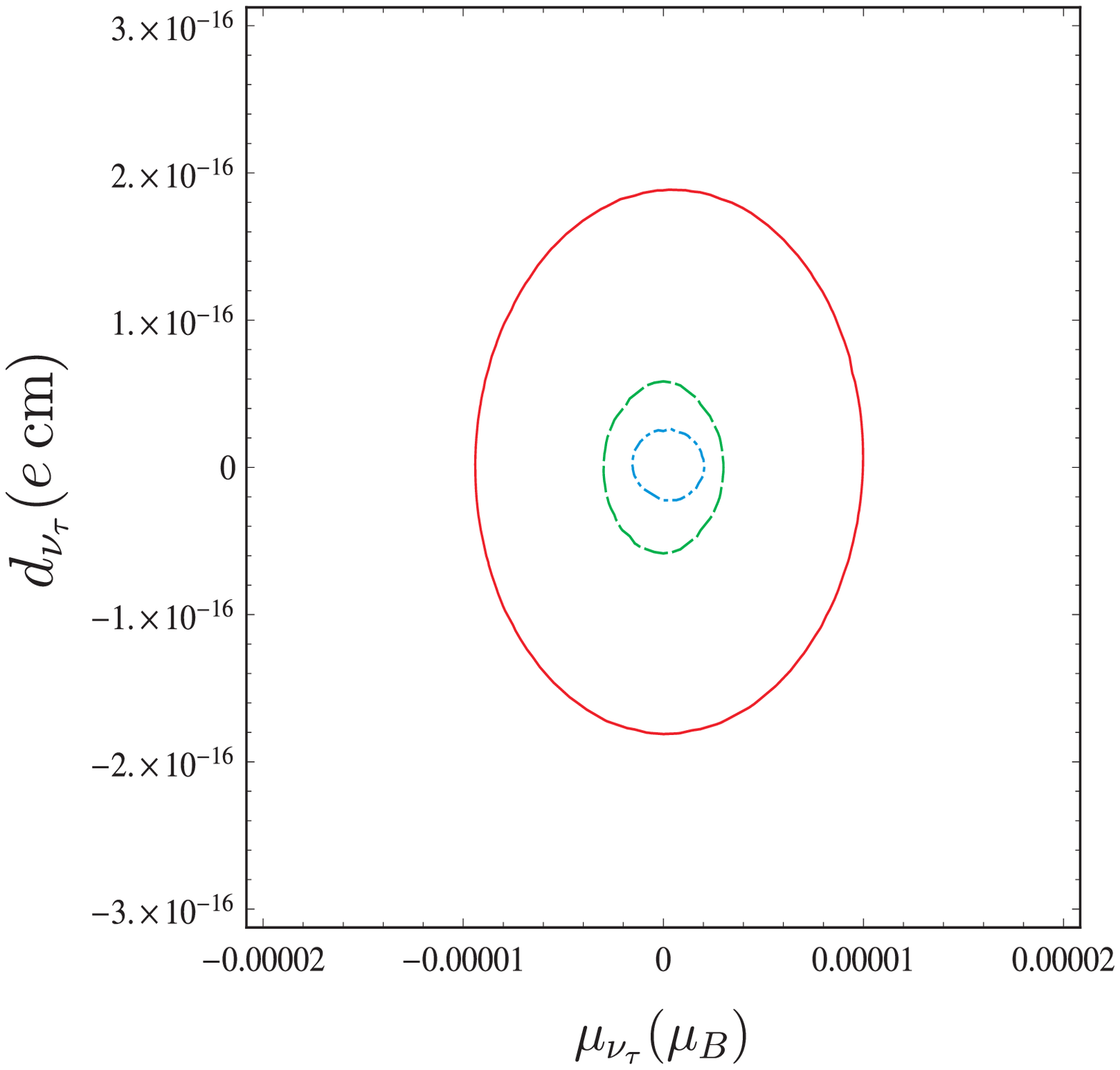}}}
\caption{\label{fig:gamma17} Limits contours at the $95\% \hspace{1mm}C. L.$ in the $\mu_{\nu_\tau}-d_{\nu_\tau}$ plane for
$e^+e^- \to e^+\gamma^* e^- \to e^+\tau \bar\nu_\tau \nu_e $. Starting from the top, the curves are for $\sqrt{s}=0.5$\hspace{1mm}$TeV$
and ${\cal L}=230$ $fb^{-1}$; $\sqrt{s}=1.5$\hspace{1mm}$TeV$ and ${\cal L}=320$ $fb^{-1}$; $\sqrt{s}=3$\hspace{1mm}$TeV$
and ${\cal L}=590$ $fb^{-1}$, respectively. We have used $Q^2=2$\hspace{1mm}$GeV^2$.}
\end{figure}

\begin{figure}[t]
\centerline{\scalebox{0.77}{\includegraphics{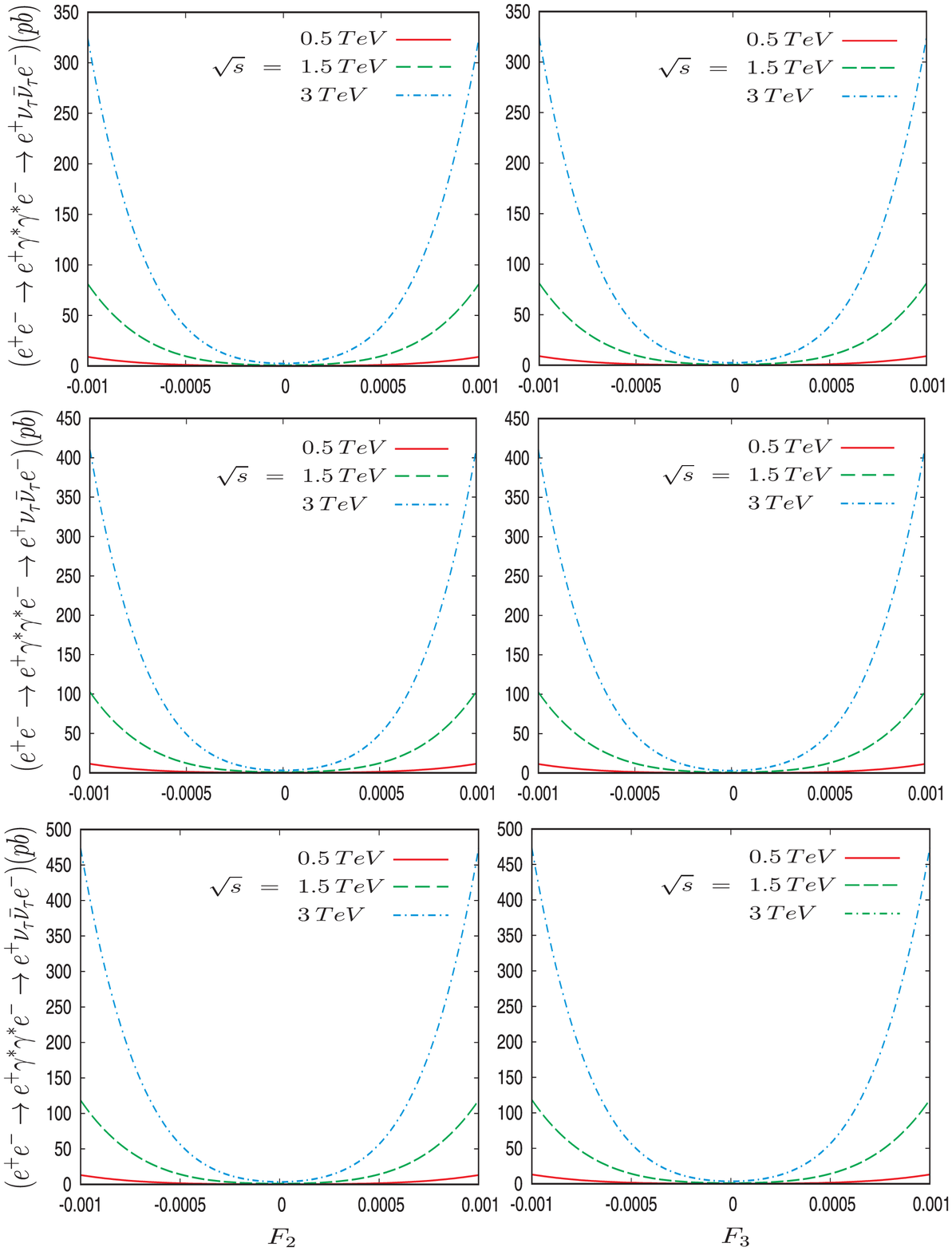}}}
\caption{ \label{fig:gamma5} The total cross section of the process
$e^+e^- \to e^+\gamma^* \gamma^* e^- \to e^+\nu_\tau \bar\nu_\tau e^- $ as a function of the
anomalous couplings $F_2$ and $F_3$ for three different center-of-mass energies
$\sqrt{s}=0.5, 1.5, 3$\hspace{1mm}$TeV$ and $Q^2=2, 16, 64$\hspace{1mm}$GeV^2$,
respectively.}
\end{figure}

\begin{figure}[t]
\centerline{\scalebox{0.76}{\includegraphics{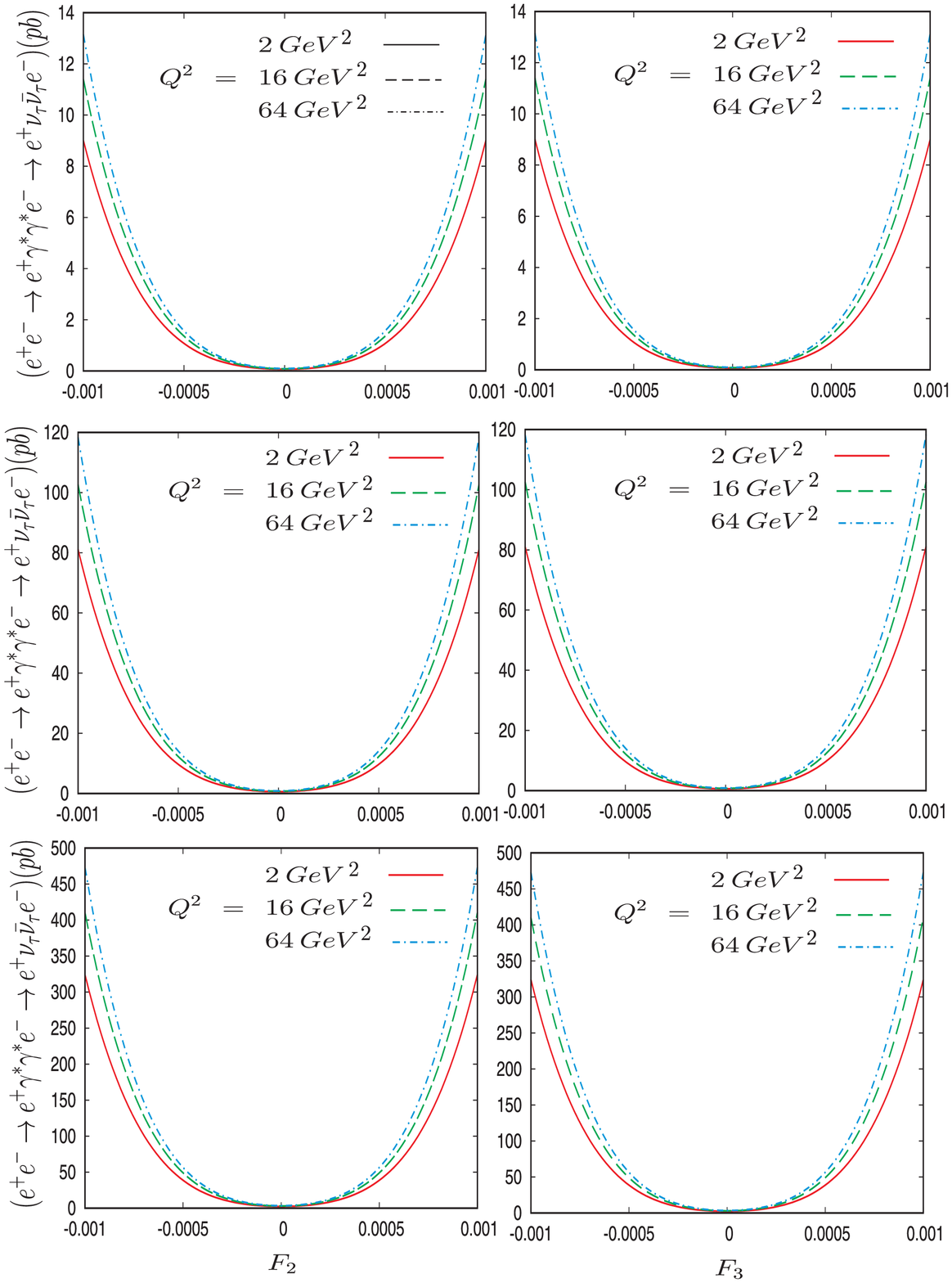}}}
\caption{ \label{fig:gamma6} The integrated total cross-section of the process
$e^+e^- \to e^+\gamma^* \gamma^* e^- \to e^+ \nu_\tau \bar \nu_{\tau} e^-$ as a function of the
anomalous couplings $F_2$ and $F_3$ for three different values of $Q^2=2, 16, 64$\hspace{1mm}$GeV^2$
and center-of-mass energies $\sqrt{s}=0.5, 1.5, 3$\hspace{1mm}$TeV$, respectively.}
\end{figure}

\begin{figure}[t]
\centerline{\scalebox{0.77}{\includegraphics{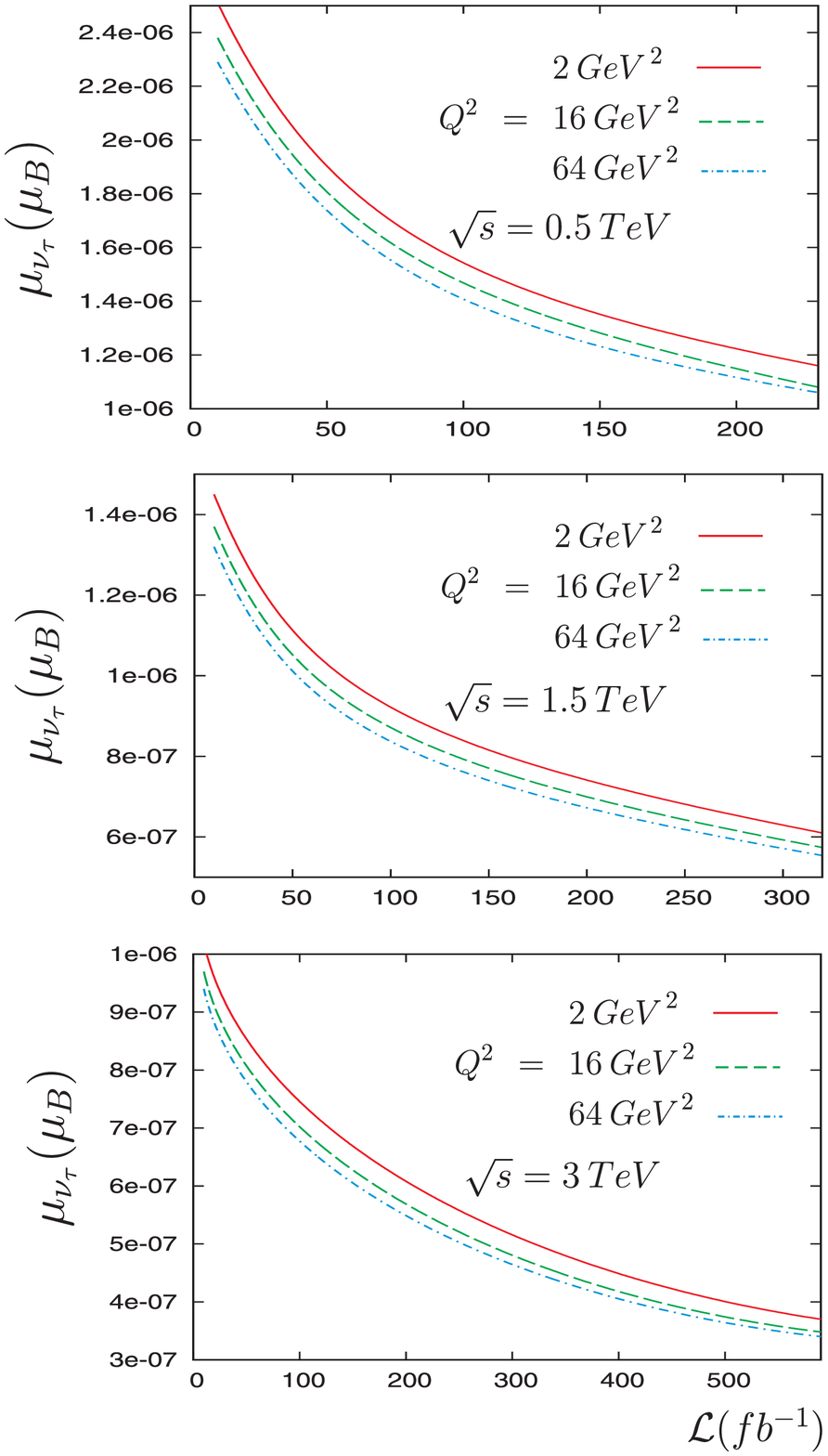}}}
\caption{\label{fig:gamma17} Dependence of the sensitivity limits at $95\% \hspace{1mm}C. L.$ for the
anomalous magnetic moment for three different values of $Q^2=2, 16, 64$\hspace{1mm}$GeV^2$ and
center-of-mass energies $\sqrt{s}=0.5, 1.5, 3$\hspace{1mm}$TeV$ in the subprocess $\gamma^* \gamma^* \to \nu_\tau \bar \nu_{\tau}$.}
\end{figure}

\begin{figure}[t]
\centerline{\scalebox{0.77}{\includegraphics{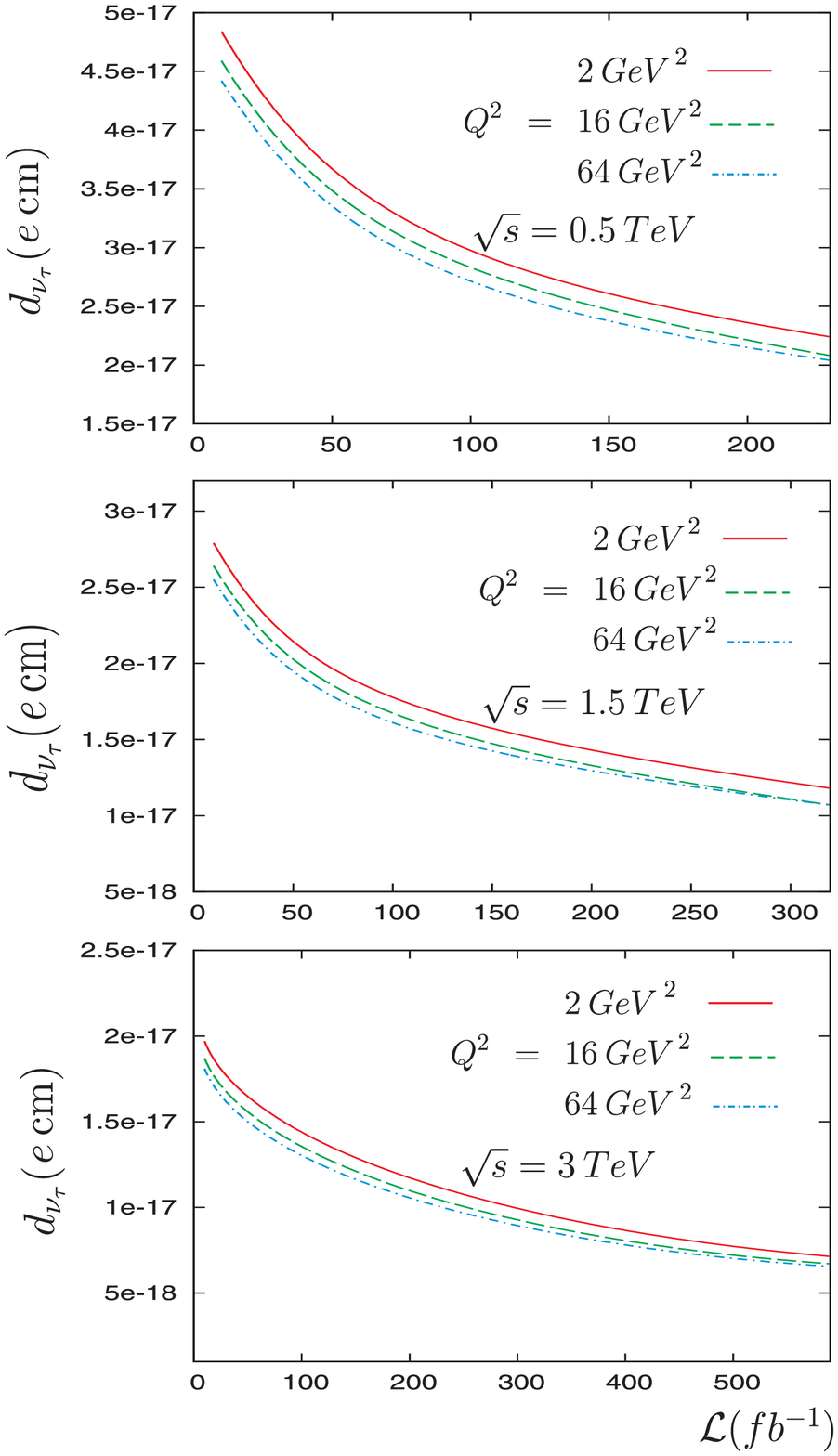}}}
\caption{\label{fig:gamma18} The same as Fig. 12 but for the electric dipole moment.}
\end{figure}

\begin{figure}[t]
\centerline{\scalebox{0.65}{\includegraphics{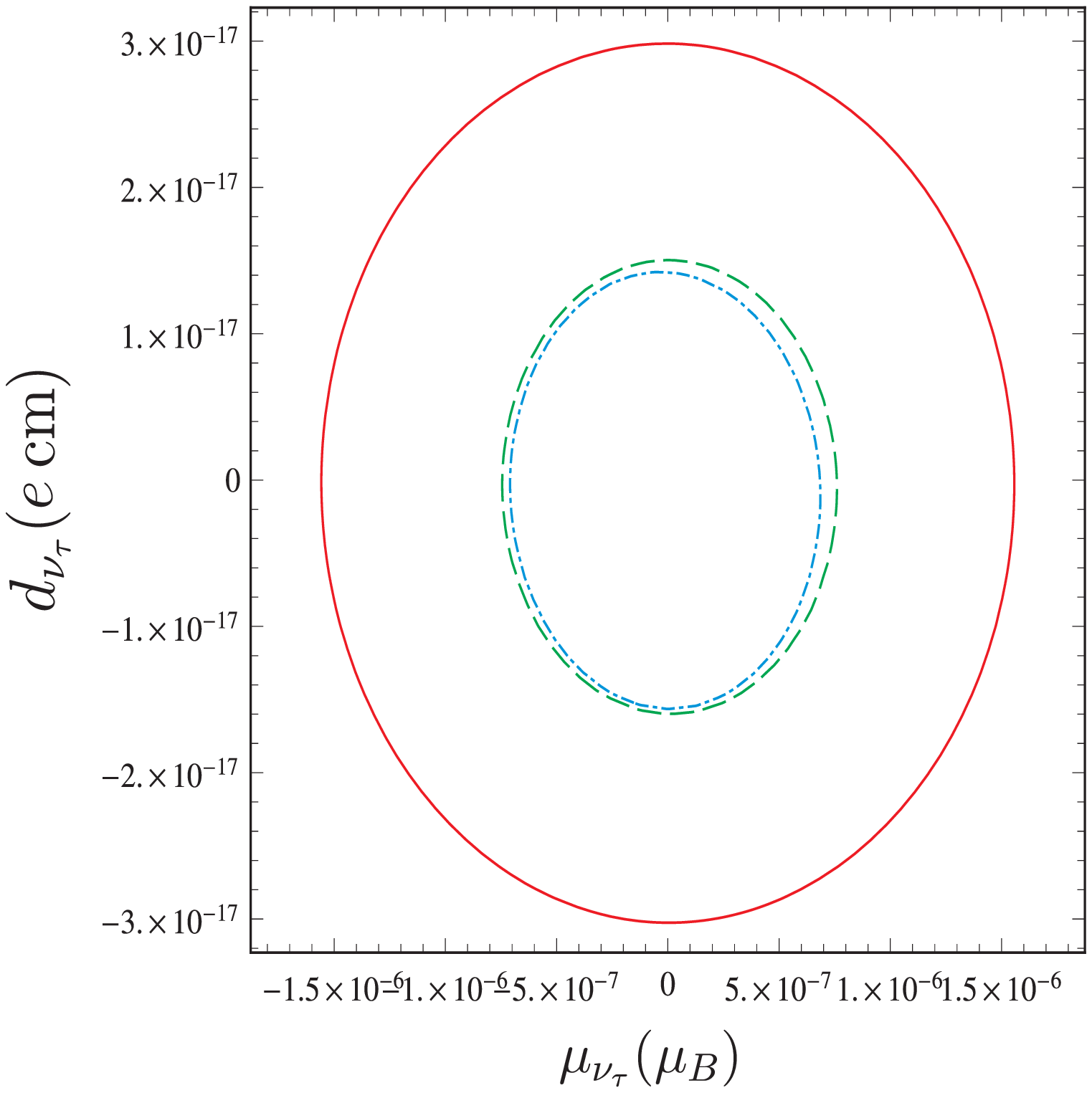}}}
\caption{\label{fig:gamma17} Limits contours at the $95\% \hspace{1mm}C. L.$ in the $\mu_{\nu_\tau}-d_{\nu_\tau}$ plane
for $e^+e^- \to e^+\gamma^* \gamma^* e^- \to e^+ \nu_\tau \bar \nu_{\tau} e^-$. Starting from the top, the curves are for
$\sqrt{s}=0.5$\hspace{1mm}$TeV$ and ${\cal L}=230$ $fb^{-1}$; $\sqrt{s}=1.5$\hspace{1mm}$TeV$ and ${\cal L}=320$ $fb^{-1}$;
$\sqrt{s}=3$\hspace{1mm}$TeV$ and ${\cal L}=590$ $fb^{-1}$, respectively. We have used $Q^2=2$\hspace{1mm}$GeV^2$.}
\end{figure}


\newpage

\begin{thebibliography}{99}

\bibitem{S.L.Glashow} S. L. Glashow,  {\it Nucl. Phys.}  {\bf 22}, 579  (1961).

\bibitem{Weinberg} S. Weinberg,  {\it Phys. Rev. Lett.} {\bf 19}, 1264  (1967).

\bibitem{Salam} A. Salam, in {\it Elementary Particle Theory},
                      Ed. N. Svartholm (Almquist and Wiskell, Stockholm, 1968) 367.

\bibitem{Fujikawa} K. Fujikawa and R. Shrock, {\it Phys. Rev. Lett.} {\bf 45}, 963 (1980).

\bibitem{Shrock} Robert E. Shrock, {\it Nucl. Phys.} {\bf B206}, 359 (1982).

\bibitem{Fukugita} M. Fukugita and T. Yanagida, {\it Physics of Neutrinos and
                   Applications to Astrophysics}, (Springer, Berlin, 2003).

\bibitem{Cisneros} A. Cisneros, {\it Astrophys. Space Sci.} {\bf 10}, 87 (1971).

\bibitem{Raffelt} G. G. Raffelt, {\it Phys Rep.} {\bf 320}, (1999) 319.

\bibitem{Bed} A. G. Bed, {\it et al.}, [GEMMA Collaboration] {\it Adv. High Energy Phys.} {\bf 2012}, (2012) 350150.

\bibitem{Auerbach} L. B. Auerbach, {\it et al.}, [LSND Collaboration] {\it Phys Rev.} {\bf D63}, (2001) 112001, hep-ex/0101039.

\bibitem{Aguila} F. del Aguila and M. Sher, {\it Phys Lett.} {\bf B252}, (1990) 116.\

\bibitem{Escribano} R. Escribano and E. Mass\'o, {\it Phys. Lett.} {\bf B395}, 369 (1997).

\bibitem{Vogel} P. Vogel and J. Engel,  {\it Phys. Rev.} {\bf D39}, 3378 (1989).

\bibitem{Bernabeu1} J. Bernabeu, {\it et al.,}  {\it Phys. Rev.} {\bf D62}, 113012 (2000).

\bibitem{Bernabeu2} J. Bernabeu, {\it et al.,}  {\it Phys. Rev. Lett.} {\bf 89}, 101802 (2000);
                   {\it Phys. Rev. Lett.} {\bf 89}, 229902 (2002).

\bibitem{Dvornikov} M. S. Dvornikov and A. I. Studenikin, {\it Jour. of Exp. and Theor. Phys.} {\bf 99}, 254 (2004).

\bibitem{Giunti} C. Giunti and A. Studenikin, {\it Phys. Atom. Nucl.} {\bf 72}, 2089 (2009).

\bibitem{Broggini} C. Broggini, C. Giunti, A. Studenikin, {\it Adv. High Energy Phys.} {\bf 2012}, (2012) 459526;
                   arXiv:1207.3980 [hep-ph] and references therein.


\bibitem{Borexino} C. Arpesella, {\it et al.}, [Borexino Collaboration], {\it Phys. Rev. Lett.} {\bf 101}, 091302 (2008).

\bibitem{DONUT} R. Schwinhorst, {\it et al.}, [DONUT Collaboration], {\it Phys. Lett.} {\bf B513}, 23 (2001).

\bibitem{A.M.Cooper} A. M. Cooper-Sarkar, {\it et al.}, [WA66 Collaboration], {\it Phys. Lett.} {\bf B280}, 153 (1992).

\bibitem{L3} M. Acciarri {\it et al.}, [ L3 Collaboration], {\it Phys. Lett.} {\bf B412}, 201 (1997).

\bibitem{Gutierrez9} A. Guti\'errez-Rodr\'iguez, {\it Int. J. Theor. Phys.} {\bf 54}, 236 (2015).

\bibitem{Gutierrez8} A. Guti\'errez-Rodr\'iguez, {\it Advances in High Energy Physics} {\bf 2014}, 491252 (2014).

\bibitem{Data2014} K. A. Olive,  {\it et al.}, [Particle Data Group], {\it Chin. Phys.}  {\bf C38}, 090001 (2014).

\bibitem{Gutierrez7} A. Guti\'errez-Rodr\'iguez, {\it Pramana Journal of Physics} {\bf 79}, 903 (2012).

\bibitem{Gutierrez6} A. Guti\'errez-Rodr\'iguez, {\it Eur. Phys. J.} {\bf C71}, 1819 (2011).

\bibitem{Aydin} C. Aydin, M. Bayar and N. Kilic, {\it Chin. Phys.} {\bf C32}, 608 (2008).

\bibitem{Gutierrez5} A. Guti\'errez-Rodr\'iguez,  {\it et al.}, {\it Phys. Rev.} {\bf D74}, 053002 (2006).

\bibitem{Perez} M. A. P\'erez, G. Tavares-Velasco and J. J. Toscano, {\it Int. J. Mod. Phys.} {\bf A19}, 159 (2004).

\bibitem{Gutierrez4} A. Guti\'errez-Rodr\'iguez,  {\it et al.}, {\it Phys. Rev.} {\bf D69}, 073008 (2004).

\bibitem{Gutierrez3} A. Guti\'errez-Rodr\'iguez,  {\it et al.}, {\it Acta Physica Slovaca}  {\bf 53}, 293 (2003).

\bibitem{Larios} F. Larios, M. A. P\'erez, G. Tavares-Velasco, {\it Phys. Lett.} {\bf B531}, 231 (2002).

\bibitem{Keiichi} K. Akama, T. Hattori and K. Katsuura, {\it Phys. Rev. Lett.}  {\bf 88}, 201601 (2002).

\bibitem{Aytekin} A. Aydemir and R. Sever, {\it Mod. Phys. Lett.} {\bf A16} 7, 457 (2001).

\bibitem{Gutierrez2} A. Guti\'errez-Rodr\'iguez, {\it et al.},  {\it Rev. Mex. de F\'{\i}s.} {\bf 45}, 249 (1999).

\bibitem{Hernandez} J. M. Hern\'andez,  {\it et al.},  {\it Phys. Rev.} {\bf D60}, 013004 (1999).

\bibitem{Maya} M. Maya, M. A. P\'erez, G. Tavares-Velasco, B. Vega,  {\it Phys. Lett.} {\bf B434}, 354 (1998).

\bibitem{Gutierrez1} A. Guti\'errez-Rodr\'iguez,  {\it et al.},  {\it Phys. Rev.} {\bf D58}, 117302 (1998).

\bibitem{DELPHI} P. Abreu, {\it et al.}, [DELPHI Collaboration],  {\it Z. Phys.} {\bf C74}, 577 (1997).

\bibitem{Gould} T. M. Gould and I. Z. Rothstein,  {\it Phys. Lett.} {\bf B333}, 545 (1994).

\bibitem{Grotch} H. Grotch and R. Robinet,  {\it Z. Phys.} {\bf C39}, 553 (1988).

\bibitem{Abe} T. Abe, {\it et al.} (Am. LC Group), arXiv:hep-ex/0106057; G. Aarons et al., (ILC Collaboration),
              arXiv: 0709.1893 [hep-ph]; J. Brau et al., (ILC Collaboration), arXiv: 0712.1950 [physics.acc-ph];
              H. Baer, T. Barklow, K. Fujii et al., arXiv:1306.6352 [hep-ph].

\bibitem{Accomando} E. Accomando, {\it et al.} (CLIC Phys. Working Group Collaboration), arXiv: hep-ph/0412251, CERN-2004-005;
                    D. Dannheim, P. Lebrun, L. Linssen et al., arXiv: 1208.1402 [hep-ex];
                    H. Abramowicz et al., (CLIC Detector and Physics Study Collaboration), arXiv:1307.5288 [hep-ph].

\bibitem{Budnev} V. M. Budnev, I. F. Ginzburg, G. V. Meledin and V. G. Serbo, {\it Phys. Rep.} {\bf 15}, 181 (1975).

\bibitem{Baur} G. Baur, {\it et al.}, {\it Phys. Rep.} {\bf 364}, 359 (2002).

\bibitem{Piotrzkowski} K. Piotrzkowski, {\it Phys. Rev.} {\bf D63}, 071502 (2001).

\bibitem{Abulencia} A. Abulencia, {\it et al.}, [CDF Collaboration],  {\it Phys. Rev. Lett.} {\bf 98}, 112001 (2007).

\bibitem{Aaltonen1} T. Aaltonen, {\it et al.}, [CDF Collaboration], {\it Phys. Rev. Lett.} {\bf 102}, 222002 (2009).

\bibitem{Aaltonen2} T. Aaltonen, {\it et al.}, [CDF Collaboration], {\it Phys. Rev. Lett.} {\bf 102}, 242001 (2009).

\bibitem{Chatrchyan1} S. Chatrchyan, {\it et al.}, [CMS Collaboration], {\it JHEP} {\bf 1201}, 052 (2012).

\bibitem{Chatrchyan2} S. Chatrchyan, {\it et al.}, [CMS Collaboration], {\it JHEP} {\bf 1211}, 080 (2012).

\bibitem{Abazov} V. M. Abazov, {\it et al.}, [D0 Collaboration], {\it Phys. Rev.} {\bf D88}, 012005 (2013).

\bibitem{Chatrchyan3} S. Chatrchyan, {\it et al.}, [CMS Collaboration], {\it JHEP} {\bf 07}, 116 (2013).

\bibitem{DELPHI1} J. Abdallah, {\it et al.}, [DELPHI Collaboration], {\it Eur. Phys. J.} {\it C35}, (2004) 159.

\bibitem{Sahin} I. Sahin, {\it Phys. Rev.} {\bf D85}, 033002 (2012).

\bibitem{Sahin1} I. Sahin and M. Koksal, JHEP {\bf 03}, 100 (2011).

\bibitem{Pukhov} Pukhov A, {\it et al}, CompHEP—a package for evaluation of Feynman diagrams and integration over
                 multiparticle phase space, Report No. INP MSU 98-41/542, arXiv:hep-ph/9908288.

\end{thebibliography}
\end{document}